\documentclass[11pt,twoside]{article}
\usepackage{fleqn,espcrc1}
\usepackage{epsfig}
\usepackage[figuresright]{rotating}

\newcommand{\AmS}{{\protect\the\textfont2
  A\kern-.1667em\lower.5ex\hbox{M}\kern-.125emS}}
\title{Electron capture rates for core collapse supernovae}

\author{J. M. Sampaio\address[ifa]{Institut for Physik og Astronomi, Aarhus Universitet, DK-8000 \AA rhus C, Denmark}
        K. Langanke\addressmark[ifa]
	G. Mart\'{\i}nez-Pinedo\address[basel]{Department f\"ur Physik und Astronomie der Universit\"at Basel, Basel, Switzerland}
	E. Kolbe\addressmark[basel]
	and
	D. J. Dean\address{Physics Division, Oak Ridge National Laboratory, Oak Ridge, TN 37831, USA}}  

\begin{document}
\maketitle
The collapse of the inner core of a massive star, towards a type II supernova explosion, starts with electron captures on nuclei. This has three important consequences: (i) it reduces the electron degeneracy pressure, accelerating the collapse; (ii) it drives the matter to more neutron-rich nuclei; and (iii) consequently, it produces large amounts of neutrinos, which can cool the star. The subsequent collapse evolution is strongly influenced by the deleptonization due to electron captures on successively more neutron-rich nuclei and free protons \cite{bethe}. Further deleptonization is blocked after neutrinos become trapped at core densities around $\rho\approx 6\times 10^{11}$ gcm$^{-3}$. The value of the lepton fraction at the point of neutrino thermalization is crucial for determining the size of the homologous core and, consequently, the location of the onset of the shock-wave. 

Current collapse simulations show a self-regulation mechanism that establishes similar electron fractions for all progenitor models, whenever electron captures on free protons dominate \cite{mathias,messer}. These simulations assume that electron captures on nuclei stop for nuclei with neutron numbers above the pf-shell closure $N=40$. Since Gamow-Teller (GT) transitions can only excite nucleons inside a major shell, it implies that they are Pauli blocked for nuclei with $N>40$ and $Z<40$. However, (i) thermal excitation can promote protons and neutrons to the gds orbitals \cite{cooper} and (ii) the residual interaction between nucleons mixes the gds orbitals with the pf shell. Both mechanisms unblock effectively the GT transitions for heavier nuclei with possible consequences on the self-regulation mechanism that remain to be investigated in future collapse simulations. 

Electron capture rates on nuclei in the mass range $45\leq A\leq65$ have been calculated recently, based on large-scale shell-model diagonalization within the full pf shell \cite{klgmp}. For heavier nuclei, direct diagonalization of the full pf+gds orbitals is not feasible with the present computer capabilities. A hybrid model has been proposed in \cite{germanium} to carry out this task. Here we follow the same procedure with a model space that includes the full pf+gds orbitals and considers an inert $^{\rm 40}$Ca core. The capture rates were calculated within the Random Phase Approximation (RPA) with partial number formalism, including allowed and forbidden transitions. The partial occupation numbers were provided as a function of temperature by Shell-Model Monte Carlo (SMMC) calculations \cite{smmc}, including an appropriate pairing+quadrupole interaction. In this way we could incorporate relevant features of nuclear structure, while simultaneously avoiding the sign problem in the SMMC calculations.     

\begin{figure}
\begin{minipage}{7.9cm}
\epsfig{file=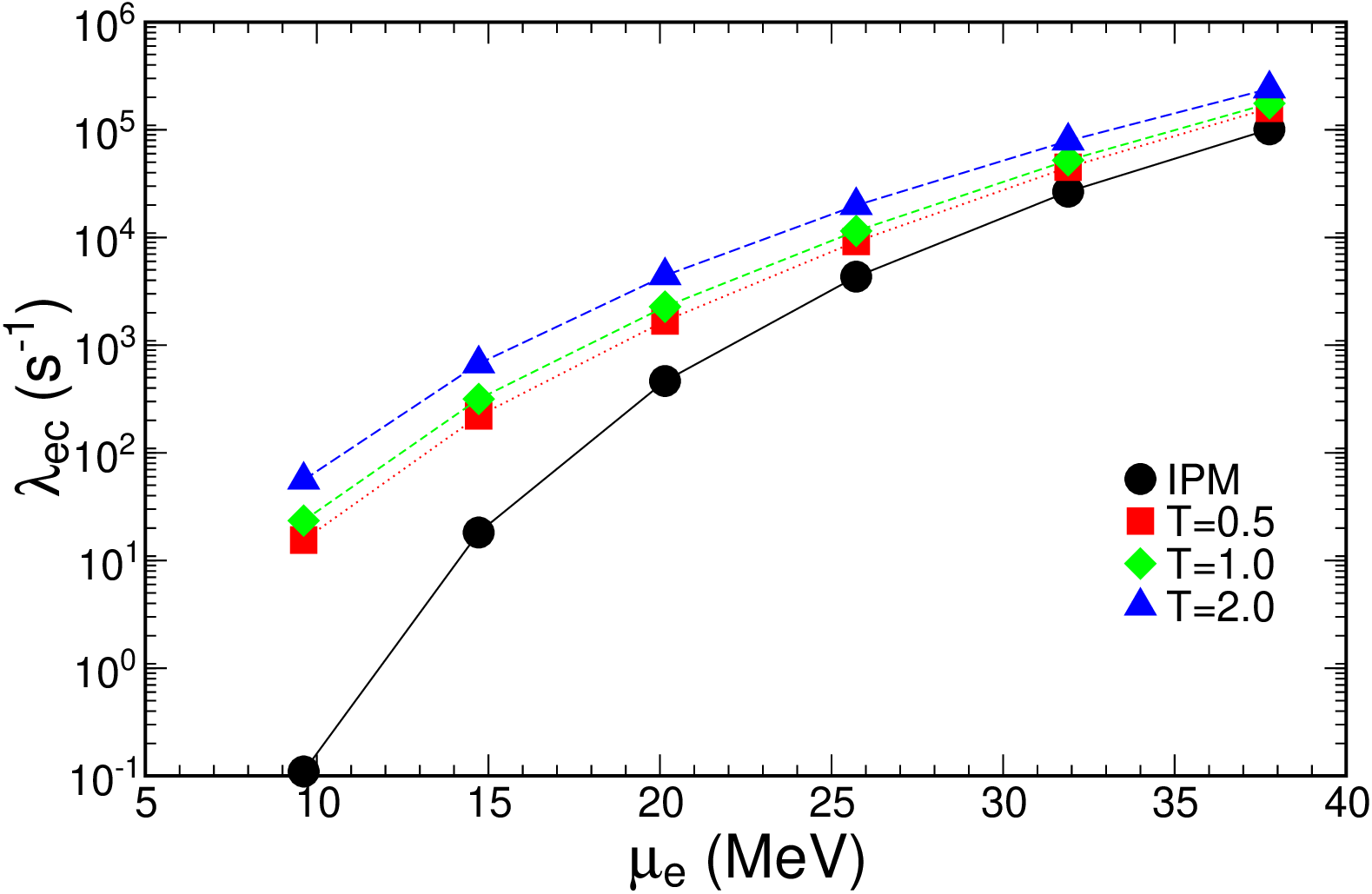,width=7.9cm}
\caption{Electron capture rates on $^{\rm 76}$Ga ($Q\approx 4.5$ MeV) as a function of the electron chemical potential. Temperatures $T$ are in MeV.}\label{ga76rates}
\end{minipage}
\hspace*{0.1cm}
\begin{minipage}{7.9cm}
\epsfig{file=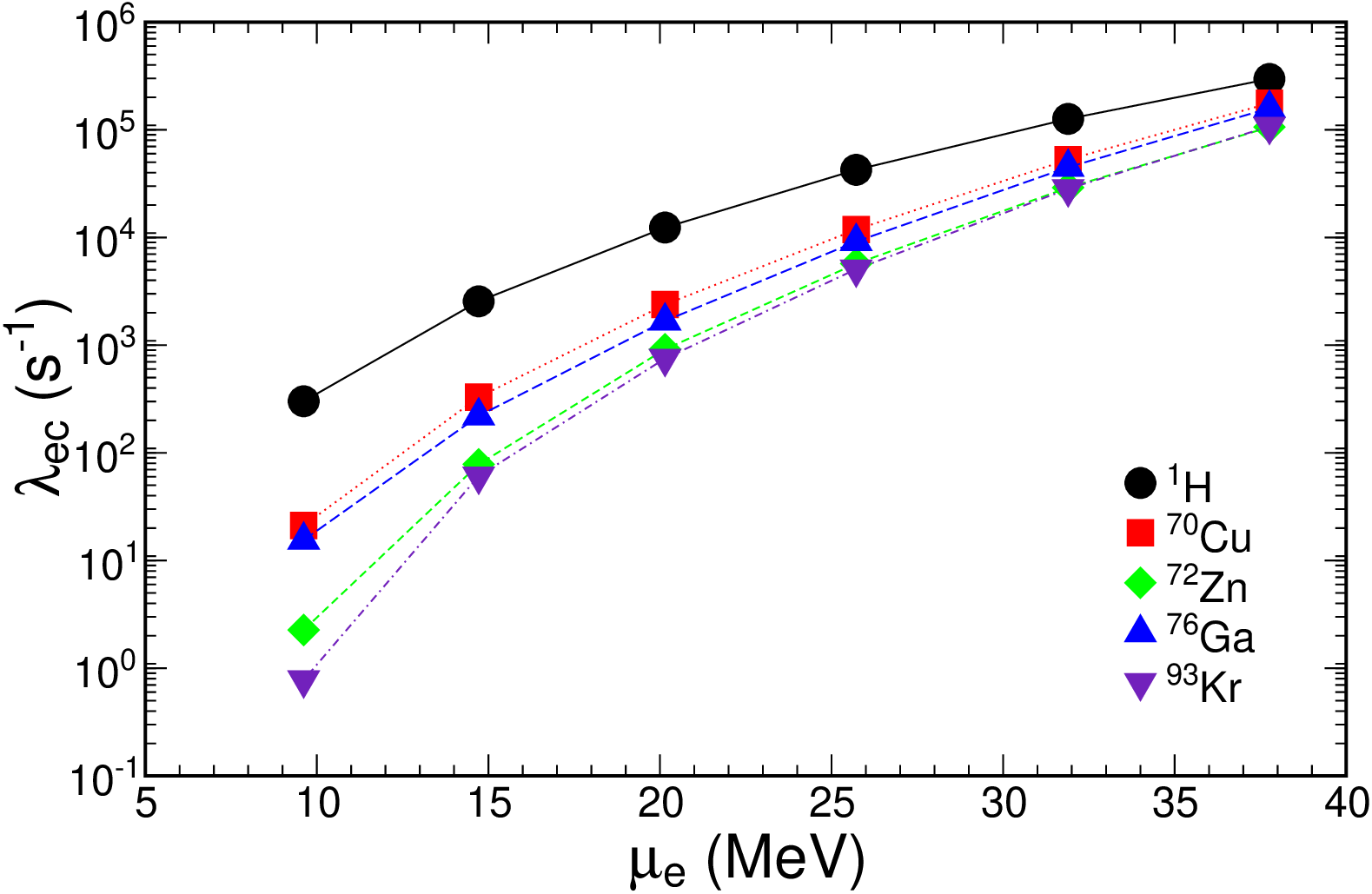,width=7.9cm}
\caption{Electron capture rates on free protons and representative nuclei during the collapse phase as a function of the electron chemical potential.}\label{rates0}
\end{minipage}
\end{figure}
Fig. \ref{ga76rates} compares the electron capture rates on $^{\rm 76}$Ga ($N=45$) within the Independent Particle Model (IPM) with the rates evaluated at three different temperatures using the SMMC/RPA model. The electron chemical potentials, $\mu_e$, were calculated following a $M=0.6M_\odot$ mass trajectory during the collapse. The competion between the electron chemical potential and the reaction Q-value determines electron capture rates on nuclei. At low densities ($\mu_e\sim Q$), allowed GT transitions dominate the captures in the SMMC/RPA model, while, in the IPM these transitions are blocked and only forbidden transitions contribute. At higher densities ($\mu_e\gg Q$), the capture rates become less sensitive to the Q-value. Furthermore, forbidden transitions contribute significantly at higher electron energies and, hence, the rates, in the IPM and in the SMMC/RPA model, converge to the same value. 

Electron capture rates on relevant nuclei above $A=65$ have been calculated for the same stellar conditions as adopted in Fig. \ref{ga76rates}. Fig. \ref{rates0} shows the electron capture rates on free protons and five representative nuclei during the collapse phase as a function of the electron chemical potential. The capture rates on free protons are larger than the capture rates on nuclei. This is especially true for low chemical potentials, where the reaction Q-value (larger for captures on nuclei than for captures on free protons) have a strong influence on the rates. At larger chemical potentials, electron captures become approximately independent of the reaction Q-value, making the capture rates on nuclei similar to the capture rates on free protons. 

\begin{figure}
\begin{minipage}{7.9cm}
\epsfig{file=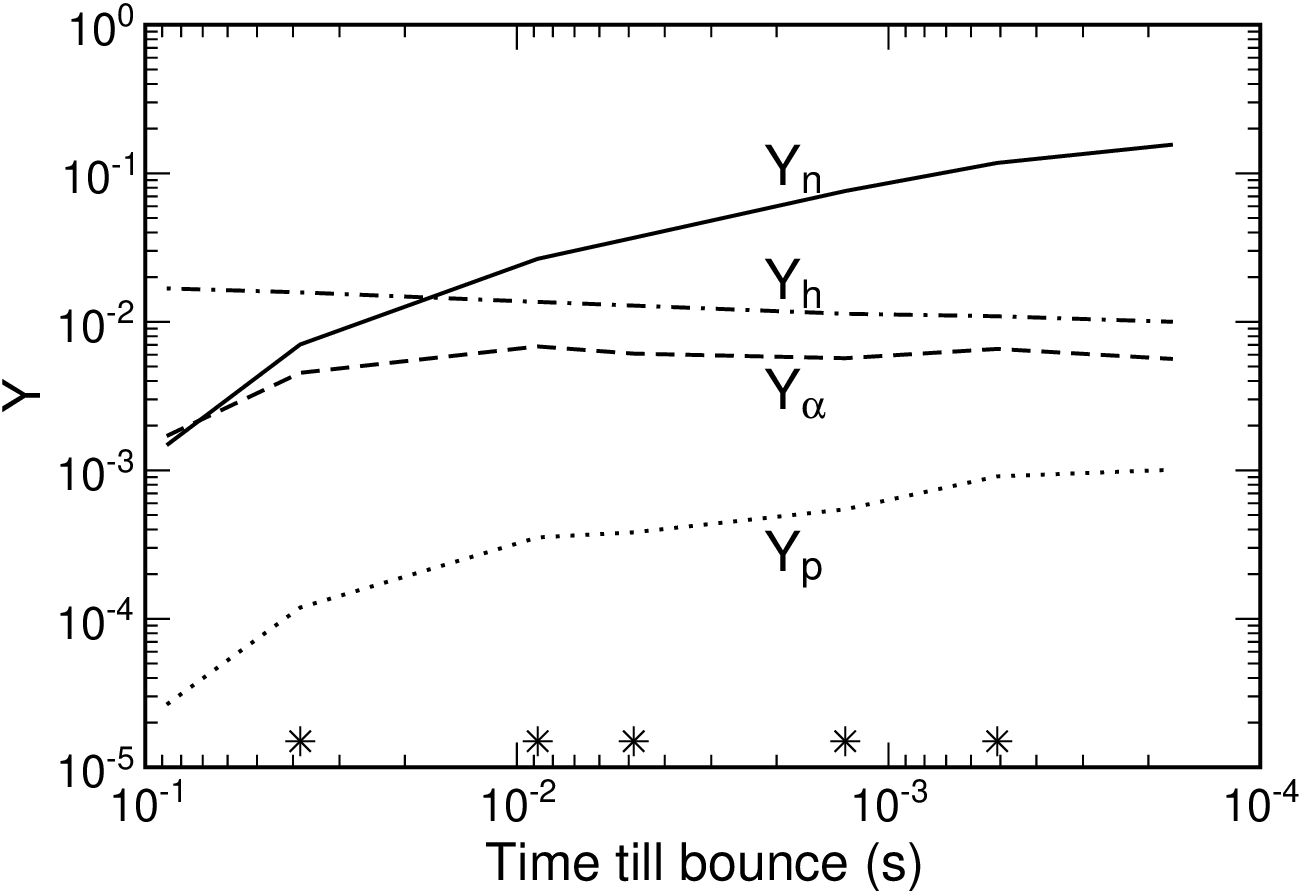,width=7.5cm}
\caption{Time evolution of the number abundances of neutrons, $Y_n$, protons, $T_p$, alpha particles, $Y_\alpha$ and heavy elements, $Y_h$, for a $M=0.6M\odot$ stellar trajectory, assuming NSE \cite{klgmp2}.}\label{nse}
\end{minipage}
\hspace*{0.1cm}
\begin{minipage}{7.9cm}
\epsfig{file=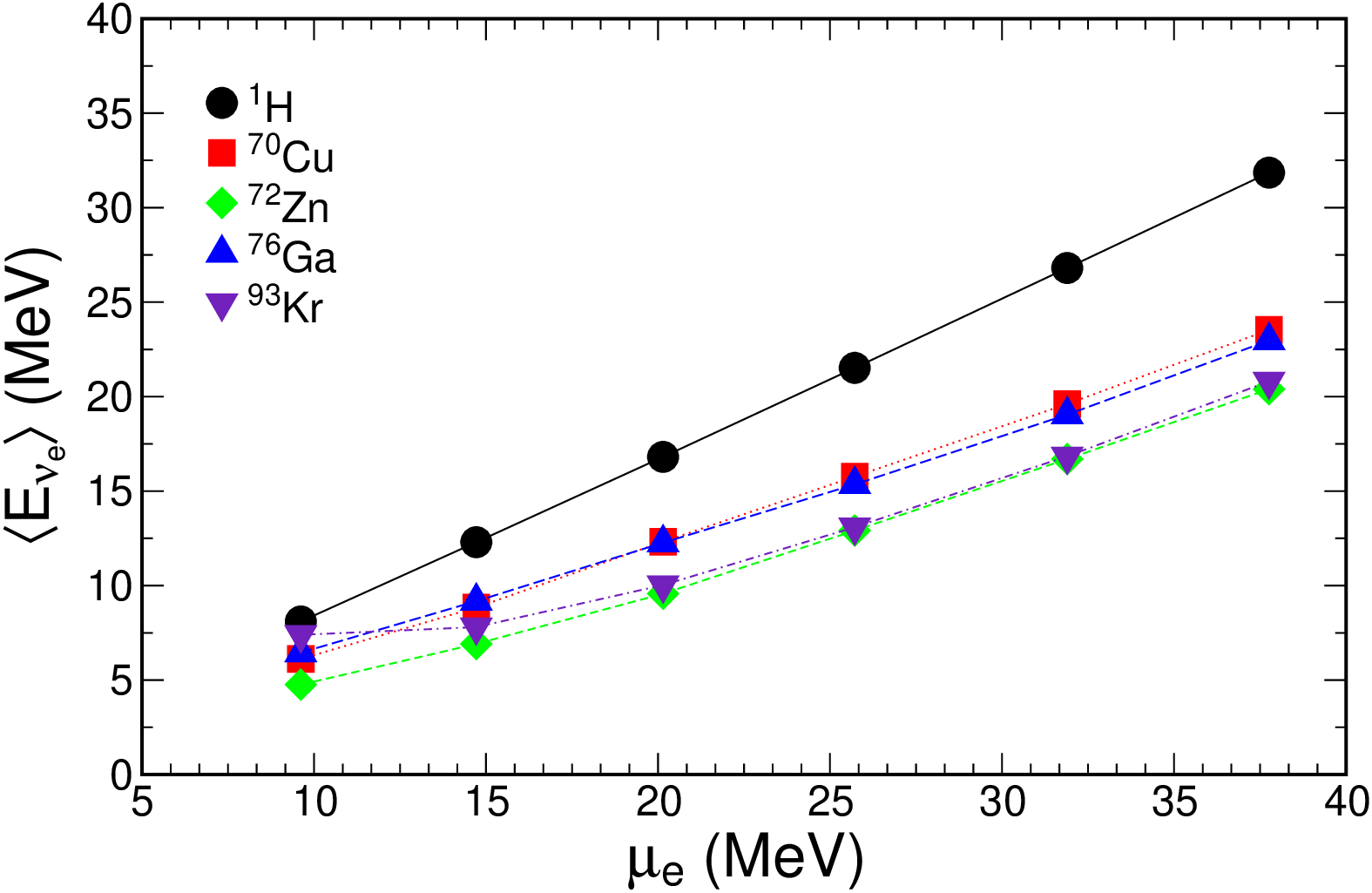,width=7.5cm}
\caption{Average neutrino energies from electron capture on free protons and representative nuclei during the collapse as a function of the electron chemical potential.}\label{neutrino}
\end{minipage}
\end{figure}

The competition between electron captures on free protons and electron captures on nuclei is determined by the product of the number abundance of a given nuclear specie, $Y$, and its capture rate, $\lambda_{\rm ec}$. In Fig. \ref{nse} the time evolution of the number abundances of neutrons, $Y_n$, protons, $Y_p$, and heavy elements, $Y_h$, is shown for the same stellar trajectory and assuming Nuclear Statistical Equilibrium (NSE). When the chemical potential is about $\mu_e\approx 10$ MeV the ratio of heavy nuclei abundance and proton abundance is of the order $Y_h/Y_p\approx 100$ (time till bounce $\approx 10^{-2}$ s). At a given stellar point, the nuclear composition is not made of a single nucleus, but rather by an ensemble of nuclei, where the most abundant nucleus is not necessarily the one with highest capture rate. For a rough estimate we assume that $^{\rm 72}$Zn, which is one of the most abundant nuclei at this stellar point, is representative of the nuclear ensemble, then we find that the ratio of the capture rate on this nucleus and on free protons balances approximately the ratio of abundances (see Fig. \ref{rates0}). Therefore, electron captures on nuclei can compete with captures on free protons. However, as the density approaches the neutrino trapping regime, the final state neutrino Pauli blocking must be taken into account. 

Fig. \ref{neutrino} shows the average energies of emitted electron neutrinos from capture on free protons and representative nuclei for the same stellar trajectory. Electron captures on nuclei produce neutrinos with energies significantly lower than neutrinos coming from electron captures on free protons. This is the combined result of the larger Q-value needed to be overcome on captures on  neutron-rich nuclei and the fact that most of the nuclear transition strength in the daughter nuclei has an excitation energy of a few MeV. Thus, there is less energy available for the emitted neutrinos on capture by nuclei than on capture on free protons. Nevertheless we should note that the capture on thermally excited states in the parent nucleus can counterbalance this effect. Since low-energy neutrinos diffuse more easily out of the stellar core, electron capture rates on nuclei can be an additional source for neutrino cooling with increasing densities and temperatures. When neutrino trapping is reached, low-energy neutrinos are Pauli blocked and one can expect that electron captures on nuclei to be more significantly hindered than electron captures on free protons. The net effect of electron captures on neutron-rich nuclei can only be determined in future collapse simulations that include these processes. 

This work was supported by the Danish Research Council and by the Swiss National Science Foundation. J. M. S. acknowledges the financial support of the Portuguese Foundation for Science and Technology.


\begin{thebibliography}{langanke}\small
\bibitem{bethe} H. A. Bethe, Rev. Mod. Phys. 62 (1990) 801.
\bibitem{mathias} M. Liebend\"orfer {\em et al}, in: Proceedings of the 11th Workshop on ``Nuclear Astrophysics, Ringberg Castle, Tegernsee, Germany (2002).
\bibitem{messer} O. E. D. Messer {\em et al}, in: Proceedings of ``Nuclei in the Cosmos XVII'', Fuji-Yoshida, Japan (2002).
\bibitem{cooper} J. Cooperstein and J. Wambach, Nucl. Phys. A 420 (1984) 591.
\bibitem{klgmp} K. Langanke and G. Mart\'{\i}nez-Pinedo, Nucl. Phys. A 673 (2000) 481. 
\bibitem{germanium} K. Langanke, E. Kolbe and D. J. Dean, Phys. Rev. C 63 (2001) 032801.
\bibitem{smmc} S. E. Koonin, D. J. Dean and K. Langanke, Phys. Rep. 278 (1997) 1. 
\bibitem{klgmp2} K. Langanke and G. Mart\'{\i}nez-Pinedo, Rev. Mod. Phys. submitted (2002).
\end{thebibliography}
\end{document}